\documentstyle[12pt]{article}
  \def\thebibliography#1{\center{\bf REFERENCES}\list
   {[\arabic{enumi}]}{\settowidth\labelwidth{[#1]}\leftmargin\labelwidth
   \advance\leftmargin\labelsep
   \usecounter{enumi}}
   \def\newblock{\hskip .11em plus .33em minus -.07em}
   \sloppy
   \sfcode`\.=1000\relax}
  
  \newcounter{cap}
  {\begin{list}{Figure \arabic{cap}\hfil}{\usecounter{cap}
  \settowidth{\labelwidth}{Figure #1}%
  \setlength{\leftmargin}{\labelwidth}%
  \addtolength{\leftmargin}{\labelsep}%
  \setlength{\parsep}{2mm plus 1mm minus 1mm}
  \setlength{\itemsep}{3mm plus 2mm minus 2mm}
  }}%
  {\end{list}}
  {\begin{list}{}{\settowidth{\labelwidth}{#1}%
  \setlength{\leftmargin}{\labelwidth}%
  \addtolength{\leftmargin}{\labelsep}%
  \setlength{\itemsep}{0pt plus 1pt}
  \setlength{\parsep}{0pt plus 1pt}
  \setlength{\topsep}{0pt plus 1pt}
  \setlength{\partopsep}{0pt plus 1pt}
  \setlength{\parskip}{2mm plus 1mm minus 1mm}
  }}%
  {\end{list}}

\newcommand{\dis}{\displaystyle}

\newcommand{\AmS}{{\protect\the\textfont2
  A\kern-.1667em\lower.5ex\hbox{M}\kern-.125emS}}

\input epsf

\begin{document}
\parindent 1.3cm
\thispagestyle{empty}   
\vspace*{-2cm}
\noindent
\hspace*{10cm}
UG--FT--37/94 \\
\hspace*{10cm}
June 1994 \\ 

\begin{center}
\begin{bf}
\noindent
NON-STANDARD TAU PAIR PRODUCTION IN TWO PHOTON COLLISIONS AT LEP II AND BEYOND
\footnote{This work was partially supported by the European Union
                under contract CHRX-CT92-0004 and by CICYT under contract 
                AEN93-0615.}
\vspace{0.3cm}  \end{bf}

\noindent

Fernando Cornet and Jos\'e I. Illana\\
Depto. de F{\'\i}sica Te\'orica y del Cosmos, \\ 
Univ. de Granada, 18071 Granada, Spain
\end{center}
\vspace{3cm}

\begin{center}
\begin{minipage}{12cm}
\begin{center}
ABSTRACT	 \\
\end{center}

We study the sensitivity of LEP II and NLC, via two photon collisions,
to the effects of anomalous $\tau \overline{\tau} \gamma$ couplings.
We also discuss some CP-odd observables that can be useful to disentangle
the contributions from the anomalous couplings.

\end{minipage} 
\end{center}

\newpage

The present bounds on the anomalous magnetic and electric dipole moments
of the $e$ and $\mu$ \cite{ELECTRON,MUON}
are much stronger than the ones for the $\tau$ \cite{SILVERMAN,PACO,LOHMAN}. 
This is
particularly unfortunate since larger deviations from the Standard Model
values are expected for the $\tau$ than for the other leptons. An example
is provided by Weinberg-type models \cite{WEINBERG}, where the electric dipole
moment is generated via neutral spin $0$ bosons coupled to the leptons. The 
obtained electric dipole moment is proportional to the third power
of the lepton mass. In composite models, one would also expect larger 
effects for the tau than for the rest of the leptons. 

The advantages of studying
these anomalous couplings in two photon collisions are twofold. First, from the 
theoretical point of view it is a very clean process, since there is no
contribution from the $Z$ boson and the photons are almost real, avoiding
any possible, unknown form-factor effects. Second, the measurements are 
complementary to the ones obtained in other processes, e.g. $e^+ e^-$ 
annihilations.
The problem, however, is that very high energy $e^\pm$ beams, compared
with the $\tau$ mass, are required in order to have an $e^+ e^- \to
e^+ e^- \tau^+ \tau^-$ cross-section large enough to allow for a detailed
study. But this is just the case for LEP II and more energetic $e^+ e^-$
colliders!.
A similar analysis for heavy ion colliders has been done in \cite{NOS}

The most general form of the electromagnetic $\tau \overline{\tau} \gamma$ 
vertex compatible with Lorentz invariance and hermiticity \cite{IZ} is given
by
\begin{equation}
  \label{VERTEX}
-ie\overline{u}(p^\prime)  ( F_1 (q^2) \gamma^\mu
      + i F_2(q^2) \sigma^{\mu \nu} \dis{q_\nu \over 2 m_\tau}    
     + F_3(q^2) \gamma_5 \sigma^{\mu \nu} \dis{q_\nu \over 2 m_\tau} 
                    ) u(p) \epsilon_\mu (q),
\end{equation}
where $ \epsilon_\mu (q) $ is the polarization vector of the photon with
momentum $q$, $F_1(q^2)$ is related to the electric charge, $ e_\tau = e F_1(0)$,
and $F_{2,3}$ are the form factors related to
the magnetic and electric dipole moments, respectively, through
\begin{equation}
  \label{MOMENTS}
 \mu_\tau = {e (1+F_2(0)) \over 2 m_\tau}  \quad ; \quad
  d_\tau = {e F_3(0) \over 2 m_\tau}.
\end{equation}
In the Standard Model at tree level, $F_1(q^2) =1$ and $F_2(q^2) = F_3(q^2) =0$.
It should be noted that the $F_2$ term behaves under C and P like the 
Standard Model one, while the $F_3$ term violates CP.

The most stringent bounds on $F_2$ and $F_3$ come from the study of the
angular distribution in $e^+ e^- \to \tau^+ \tau^-$ at PETRA:
\begin{equation}
   \label{BOUNDS}
\begin{array}{lc}
|F_2| \leq 0.014 & {\hbox{\cite{SILVERMAN}}} \\
~~               &  ~~                        \\
|F_3| \leq 0.025 & {\hbox{\cite{PACO}}} 
\end{array}
\end{equation}
These bounds, however, neglect the effects of the form factors from
$q^2=0$ to $\sim 1.5 \times 10^3 \; GeV^2$, where the measurents were taken. 
A way to
avoid this problem at LEP was proposed in Ref. \cite{GRIFOLS}. Instead of 
looking at deviations from the Standard Model in tau pair production, one
should study $e^+ e^- \to \tau^+ \tau^- \gamma$. Using this method the
bound obtained is \cite{LOHMAN}
\begin{equation}
   \label{BOUNDLEP}
|F_2(0)| \; , \; |F_3(0)| \leq 0.23.
\end{equation}

The inclusion of the new terms in Eq. \ref{VERTEX} leads to unitarity 
violations leading to an enhancement in the cross-section for large $\tau$-pair
invariant masses. However, due to the effective $\gamma \gamma$ luminosity,
the cross-sections 
in the Standard Model and for reasonably small
values of $F_2$ and $F_3$
are dominated by the production
of $\tau$-pairs with low invariant masses (this is shown for the Standard
Model at LEP II in Fig. 1). We can, thus, neglect the effects introduced by the
unknown unitarization procedure.

\begin{figure}
\setlength{\unitlength}{1cm}
\begin{picture}(12.,8.)
\epsfxsize=12cm
\put(0,-5.5){\epsfbox{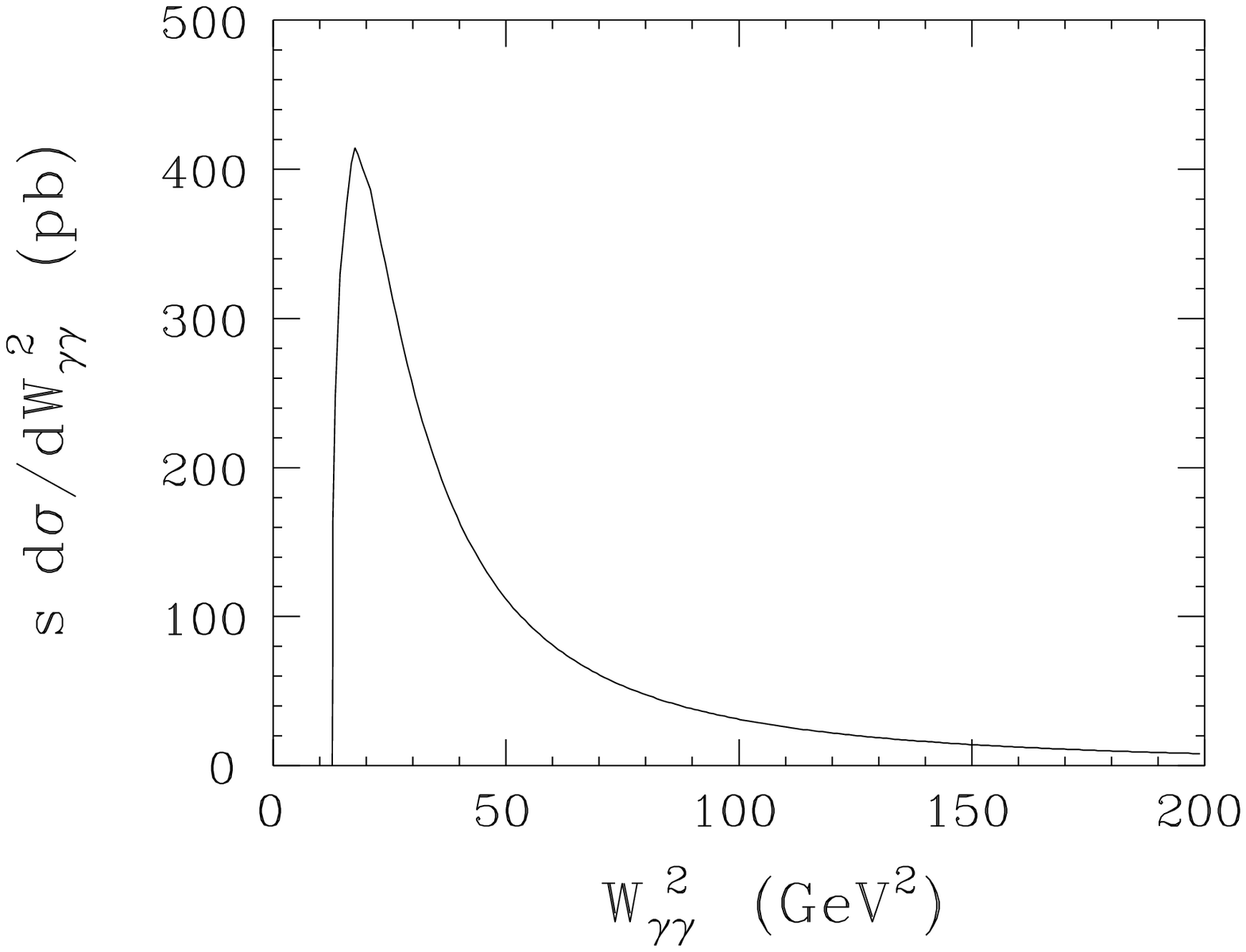}}
\end{picture}
\caption{Standard Model cross-section as a function of the two
photon invariant mass for LEP II.}
\end{figure}

The Standard Model total cross-sections are  $0.47 \; pb$ at LEP II 
and $0.792 \; pb$ at NLC. 
Assuming the integrated luminosities to be $500 \; pb^{-1}$ and
$10 \; fb^{-1}$ one expects a total amount of $235$ and $7920$ $\tau$-pairs,
respectively. In Figs. 2 and 3 we show the dependence of the total cross-section
with $F_2$ for LEP II and NLC, respectively. 
Since the relevant values of $F_2$ are small, the dependence
of the cross-section on $F_2$ can be considered as linear with a very good 
approximation:
\begin{equation}
   \label{F2SIGMA}
\begin{array}{ll}
\sigma = (0.47 + 1.55 F_2) \; pb & \qquad \hbox{LEP II} \\
~ ~ ~ ~                         &   ~ ~ ~ ~            \\
\sigma = (0.792 + 2.167 F_2) \; pb & \qquad \hbox{NLC}. \\
\end{array}
\end{equation}
This approximation is extremely good for NLC and better than a $4 \%$ for
LEP II in the range of $F_2$ values covered in the figures. The Standard
Model cross-section is obtained for $F_2 = 0$ and the expected statistical
errors for the assumed luminosities are shown with the dash lines. In this way
we see that the bounds $F_2 \leq 0.021$ and $0.004$ can be obtained at LEP II
and NLC, respectively. These bounds, however, have been obtained assuming
that all the produced $\tau$-pairs will be identified. This is certainly too
an optimistic assumption. We can exploit the simple behavior of the 
cross-section with respect to $F_2$, Eqs. \ref{F2SIGMA}, to express
the achievable bounds in terms of the luminosity and detection efficiency:
\begin{equation}
    \label{F2BOUNDS}
\begin{array}{ll}
F_2 \leq \dis{0.442 \; \hbox{pb}^{-1/2} \over 
                                   \sqrt{L\epsilon}} & \qquad \hbox{LEP II} \\
~ ~ ~ ~                         &   ~ ~ ~ ~            \\
F_2 \leq \dis{0.410 \; \hbox{pb}^{-1/2} \over  
                                   \sqrt{L\epsilon}} & \qquad \hbox{NLC}, \\
\end{array}
\end{equation}
where $\epsilon$ is the fraction of identified $\tau$-pairs and $L$ is the
integrated luminosity expressed in inverse picobarns. 
Assuming the nominal luminosity and a more realistic
situation, where $25 \%$ of the $\tau$-pairs are identified, one gets
$F_2 \leq 0.04$ and $\leq 0.008$ from LEP II and NLC, respectively. Comparing
with Eq. \ref{BOUNDS} it is clear that the bounds that can be obtained at
NLC are much more stringent than the present ones. At LEP II, however, one
can certainly improve the bounds obtained at LEP, but not the ones from
PETRA, although one should remember here that they are obtained at 
different values of $q^2$. 

\begin{figure}
\setlength{\unitlength}{1cm}
\begin{picture}(12.,8.)
\epsfxsize=12cm
\put(0,-5.5){\epsfbox{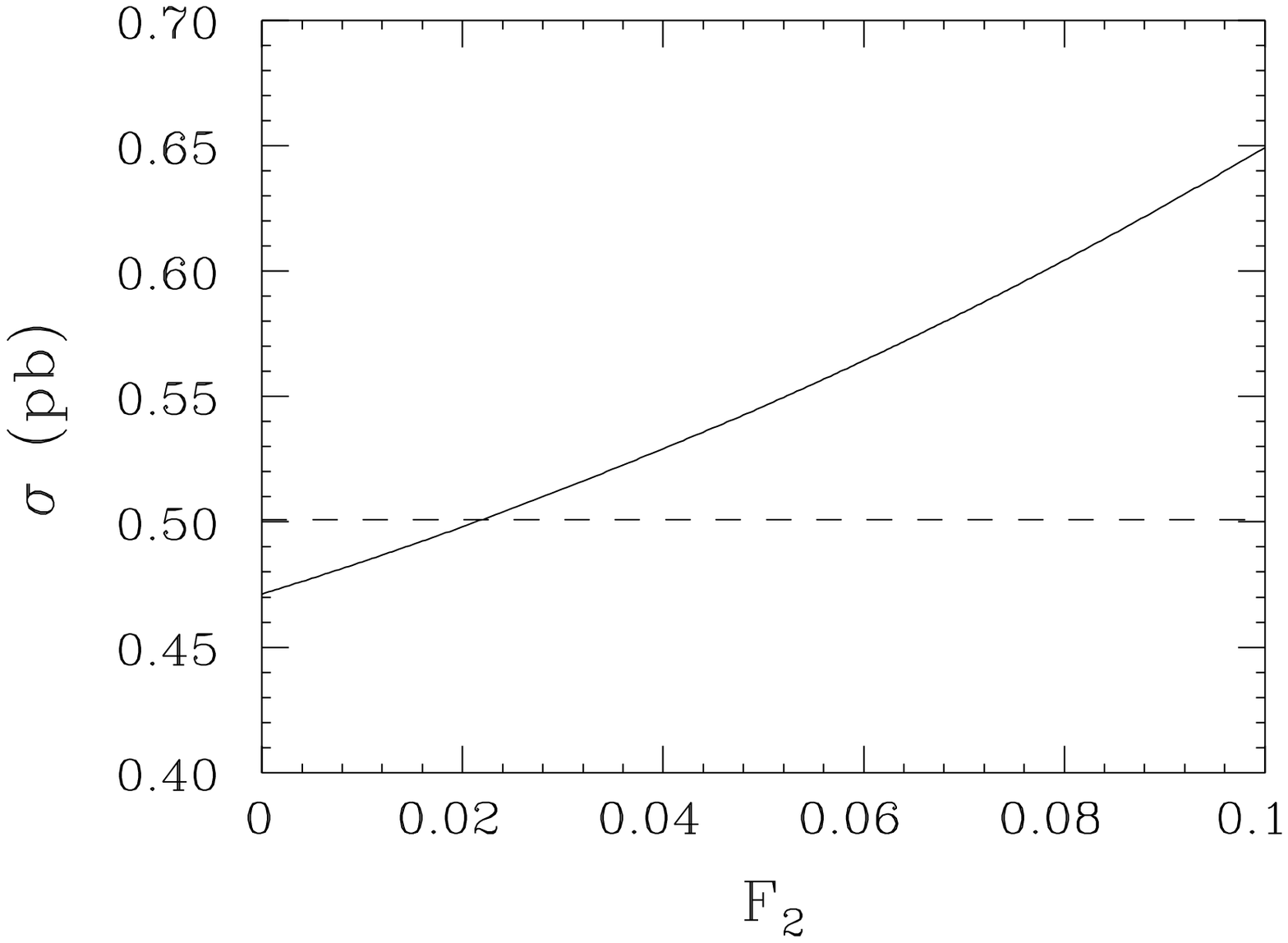}}
\end{picture}
\caption{Total cross-section for $e^+e^- \to e^+e^-\tau^+\tau^-$
as a function $F_2$ at LEP II. The dash line corresponds to
one standard deviation from the Standard Model value.}
\end{figure}
\begin{figure}
\setlength{\unitlength}{1cm}
\begin{picture}(12.,8.)
\epsfxsize=12cm
\put(0,-5.5){\epsfbox{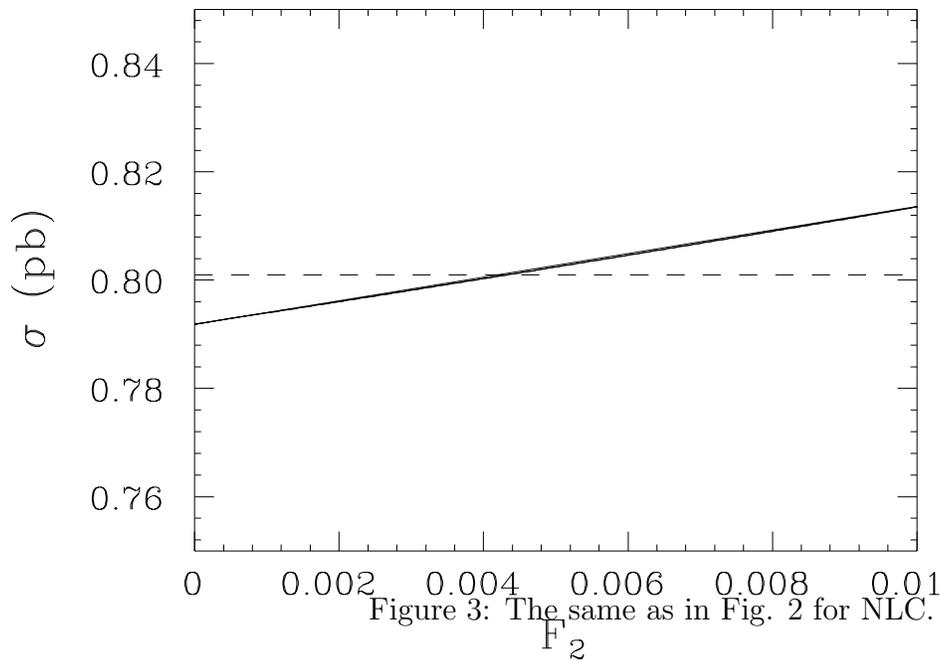}}
\end{picture}
\caption{The same as in Fig. 2 for NLC.}
\end{figure}

 The dependence of the cross-section with $F_3$ is shown in Figs. 4 and 5.
Due to the CP-violating nature of this term, the interference with the
Standard Model cancels and the dominant correction becomes $O(F_3^2)$. The
sensitivity to the new parameter will, thus, be weaker than in the case of
$F_2$. Assuming that all the produced $\tau$-pairs are identified, we
obtain, from Figs. 4 and 5, $F_3 \leq 0.08$ and $0.03$ for LEP II and NLC.
We can again obtain simple expressions for the bounds in terms of the
luminosity and the detection efficiency:
\begin{equation}
    \label{F3BOUNDS}
\begin{array}{ll}
F_3^2 \leq \dis{0.151  \; \hbox{pb}^{-1/2}  \over
                                   \sqrt{L\epsilon}} & \qquad \hbox{LEP II} \\
~ ~ ~ ~                         &   ~ ~ ~ ~            \\
F_3^2 \leq \dis{0.093  \; \hbox{pb}^{-1/2}    \over
                                   \sqrt{L\epsilon}} & \qquad \hbox{NLC}, \\
\end{array}
\end{equation}

\begin{figure}
\setlength{\unitlength}{1cm}
\begin{picture}(12.,8.)
\epsfxsize=12cm
\put(0,-5.5){\epsfbox{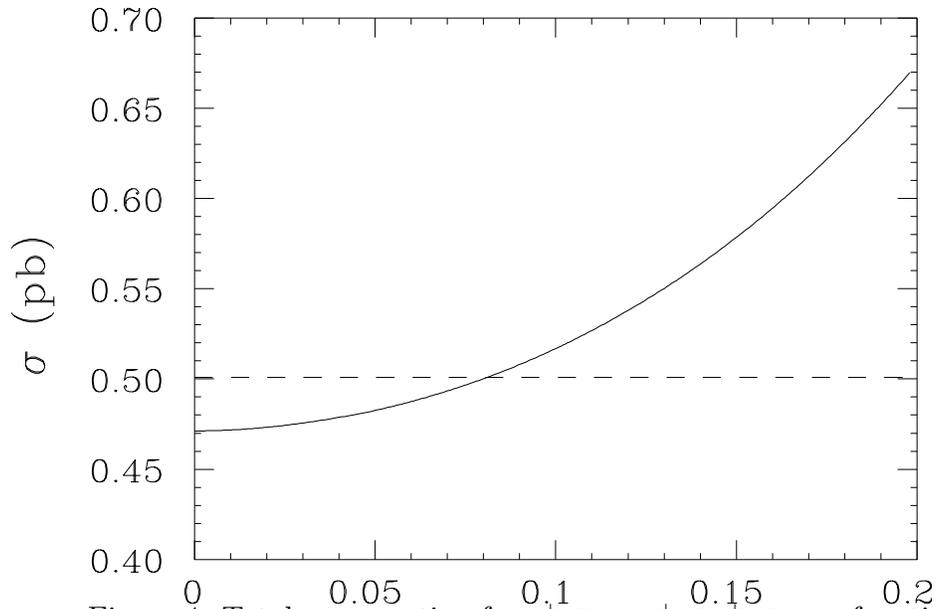}}
\end{picture}
\caption{Total cross-section for $e^+e^- \to e^+e^-\tau^+\tau^-$
 as a function $F_3$ at LEP II. The dash line corresponds to
one standard deviation from the Standard Model value.}
\end{figure}
\begin{figure}
\setlength{\unitlength}{1cm}
\begin{picture}(12.,8.)
\epsfxsize=12cm
\put(0,-5.5){\epsfbox{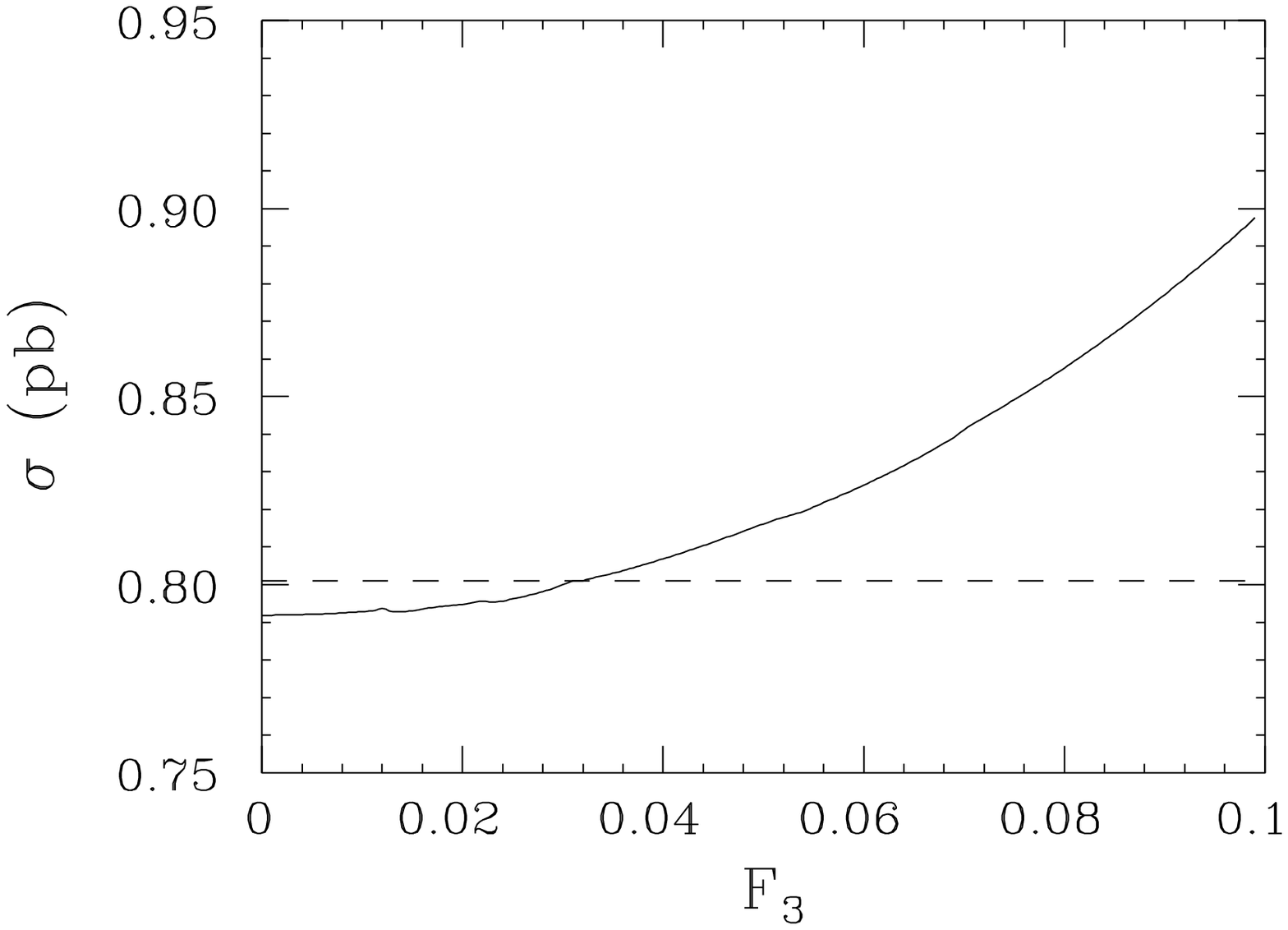}}
\end{picture}
\caption{The same as in Fig. 4 for NLC.}
\end{figure}

The bounds in Eqs. \ref{F2BOUNDS}, \ref{F3BOUNDS} are obtained from the study of
deviations in the total cross-section from the Standard Model prediction.
One could think that, similar to what it has been done at PETRA, a study of
the angular distribution would allow an improvement on the above bounds.
However, since the cross-section is dominated by low $\tau$-pair invariant
masses this is not the case for $F_2$. In the case of $F_3$ there is a small
excess of $\tau$'s produced at large angles. This effect is not large enough 
to allow a sensible improvement on the previous bounds.

We have been discussing effects of the order of a few per cent in the total
cross-section. This is of the same order as the error introduced in the
calculation when using the Weisz\"acker-Williams approximation. So, 
when comparing with real data the theoretical predictions for 
the whole $e^+ e^- \to e^+ e^- \tau^+ \tau^-$ process should be used. An 
alternative would be to normalize to the $\mu$-pair production, where the
same error in the Weisz\"acker-Williams approximation is made.

The bounds we have obtained for $F_3$ are weaker than the corresponding ones 
for $F_2$. We can try to improve these bounds exploiting the CP-violating
nature of this term. We have to look for CP-odd observables that cancel in
the Standard Model and the $F_2$ term and, thus, isolate the terms 
proportional to odd powers of $F_3$. We do not attempt a detailed study
of these CP-odd observables, rather we are going to discuss only one of them
as a showcase. The asymmetry between the cross-sections for polarized
$\tau$ and $\overline{\tau}$ production:
\begin{equation}
  \label{ASYMMETRY}
A = \dis{\sigma(s_1,s_2) - \sigma(s_2,s_1) \over 
         \sigma(s_1,s_2) + \sigma(s_2,s_1)},
\end{equation}
where the first spin is the one of the $\tau$ and the second the one of the
$\overline{\tau}$, has the required behavior under CP. We have, as an example,
taken $s_1$ and $s_2$ in the two perpendicular directions to the $\tau$ flight
direction in the respective $\tau$ or $\overline{\tau}$ rest frame.
From this asymmetry one
can find the bounds $F_3 \leq 0.05$ and $0.008$ for LEP II and NLC, 
respectively, assuming a total efficiency in the detection of the polarized
$\tau$-pairs. In terms of the luminosity and the efficiency, we have:
\begin{equation}
    \label{F3BOUNDS2}
\begin{array}{ll}
F_3 \leq \dis{1.08 \; \hbox{pb}^{-1/2} \over 
                                    \sqrt{L\epsilon}} & \qquad \hbox{LEP II} \\
~ ~ ~ ~                         &   ~ ~ ~ ~            \\
F_3 \leq \dis{0.83  \; \hbox{pb}^{-1/2}  \over
                                    \sqrt{L\epsilon}} & \qquad \hbox{NLC}. \\
\end{array}
\end{equation}
It is interesting to note that the value of the asymmetry is almost the same
for LEP II and NLC. The larger sensitivity at NLC is basically due to the better
luminosity that allows a better determination of the asymmetry.

We have discussed the possibilty of studying the anomalous electromagnetic
couplings of the $\tau$ at LEP II and NLC in $\gamma \gamma$ collisions. The
main effects of the $F_2$ term appear in deviations of the total cross-section 
from the Standard Model predictions. The bounds obtained at LEP can certainly
be improved at LEP II, but not the ones from PETRA. These bounds can only
be improved in a sensitive way at NLC. With respect to the $F_3$ term, the
sensitivity is much weaker in both colliders and one should perform more 
elaborated studies, such as CP-odd observables, in order to improve the
present bounds from PETRA.

One of us (F.C.) thanks W. Lohman for providing the bounds from LEP 
(Ref. \cite{LOHMAN}) and the
organizers of the workshop for creating a very pleasant and fruitful atmosphere
during this workshop.

\end{document}